\UseRawInputEncoding 
\pdfoutput=1
\documentclass[12pt]{article}
\usepackage{amsmath}

\usepackage{amssymb}
\usepackage{graphicx}
\usepackage{adjustbox}
\usepackage{multirow}
\usepackage{booktabs,caption}
\usepackage[flushleft]{threeparttable}
\usepackage{float}
\usepackage{xcolor}
\newcommand{\be}{\begin{equation}}
\newcommand{\ee}{\end{equation}}
\newcommand{\bea}{\begin{eqnarray}}
\newcommand{\eea}{\end{eqnarray}}

\catcode`@=12

\begin{document}
\thispagestyle{empty}
\begin{center}
{\Large\bf
{Probing Interacting Dark Energy and Scattering of Baryons with Dark Matter
in Light of EDGES 21cm Signal}}\\
\vspace{1cm}
{{\bf Upala Mukhopadhyay}\footnote{email: upala.mukhopadhyay@saha.ac.in},
{\bf Debasish Majumdar}\footnote{email: debasish.majumdar@saha.ac.in},
{\bf Kanan K. Datta}\footnote{email: kanan.physics@presiuniv.ac.in}}\\
\vspace{0.25cm}
{\normalsize \it $^{1,2}$Astroparticle Physics and Cosmology Division,}\\
{\normalsize \it Saha Institute of Nuclear Physics, HBNI,}\\ 
{\normalsize \it 1/AF Bidhannagar, Kolkata 700064, India}\\
{\normalsize \it $^3$ Department of Physics, Presidency University,}\\
{\normalsize \it 86/1 College Street, Kolkata, 700073, India}\\
\vspace{1cm}
\end{center}
\begin{abstract}
The EDGES experiment has observed an excess trough ($-500^{+200}_{-500}$ mK) in the brightness temperature $T_{21}$ of the 21cm absorption line of neutral Hydrogen atom (HI) from the era of cosmic dawn ($z \simeq 17.2$). We consider possible interaction of Dark Matter and Dark Energy fluid along with the cooling off of the baryon matter by its collision with Dark Matter to explain the observed excess trough of $T_{21}$. We make use of three different Dark Matter-Dark Energy (DM-DE) interaction models to test the viability of those models in explaining the EDGES results. The evolution of Hubble parameter is modified by DM-DE interactions and this is also addressed in this work. This in turn influences the optical depth of HI 21cm as well as the baryon temperature and thus effects the $T_{21}$ brightness temperature. In addition we also find that the DM-DE interaction enables us to explore Dark Matter with varied mass regimes and their viabilities in terms of satisfying the EDGES result.
\end{abstract}
\newpage
\section{Introduction} \label{intro}
The hyperfine transition in ground state Hydrogen atom that emits 
radiation with frequency about 1.42 GHz or in the form of 21cm hyperfine 
line could 
be an effective probe to map the Cosmos in general and the cosmic processes
during the Cosmic dark ages and the epoch of reionisation in particular
since it is the Hydrogen (around $\sim 75$ \% and helium around 25\%) that
overwhelmingly dominates the known matter of the Universe.  
The 21cm line (emission or absorption) results from the transition 
between the energy levels of 
two spin states in neutral Hydrogen at it's ground state (HI), when the 
spin directions of proton and electron in the Hydrogen atom change from 
parallel (triplet state, $S=1$) to antiparallel ($S=0$) and vice versa.    

The spin temperature $T_s$ of the 21cm transition is defined in terms 
of the populations $n_1$ and $n_0$ respectively of triplet states 
and singlet states. 
Considering 
a Boltzmann distribution one can write $n_1/n_0 = g_1/g_0 \exp(-h\nu/kT_s)$.
With $g_1 (= 3)$ and $g_0 (=1)$ being respectively the statistical weights 
of the triplet and singlet spin states, 
$k$ the Boltzmann constant and $\nu$, the  frequency ($\sim 1.42$
GHz) of the hyperfine line, one obtains, $n_1/n_0 = 3 \exp(-T_*/T_s)
\simeq 3(1 - T_*/T_s)$, $T_* = E_{\rm 21cm} = h\nu/k = 0.068$ K.
The 21cm Hydrogen absorption line which depends on the difference between 
this spin temperature $T_s$ and the background temperature is an effective
probe to understand the Universe.

Although the era of $z=1100$ is considered to be the recombination
era (the epoch of CMB last scattering) when CMB decouples from 
matter (and free stream), the large number of photons involved 
prevents such a scenario in totality as Compton scattering keeps
the matter coupled to the radiation. Thus even though the matter
temperature $T_m$ after decoupling should have cooled faster as 
$T_m \sim (1 + z)^2$ than the CMB radiation ($T_\gamma \sim (1+z)$),
this was not the case till the time scale of the Compton heating
supersedes the Hubble time. This happens at around $z\sim 200$ when 
matter actually decouples from radiation and cool adiabatically and
the matter temperature $T_m$ should follow $T_m \sim (1 + z)^2$. 
On the other hand the spin temperature $T_s$ which is expected to couple 
to CMB decouples at $z\sim20$ as the first star appears. The UV 
radiation that was emitted by those first stars initiates triplet 
to singlet transition through Wouthuysen-Field 
effect and the spin temperature moves towards the 
matter temperature $T_m$.  The combined effects of $T_s$, $T_m$ and 
the background CMB, is expressed as the brightness temperature of 
21cm Hydrogen absorption line $T_{21}$ with respect to the background 
temperature. This can be expressed as 
\begin{eqnarray}
T_{21} &=& \frac {T_s - T_\gamma} {1+z} (1 - e^{-\tau}) \simeq 
\frac{T_s - T_\gamma} {1 + z} \tau\,\,, \label{T21_1}
\end{eqnarray}
where $\tau$ is the optical depth given by \cite{J.R. Protchard} 
\begin{eqnarray}
\tau &=& \int ds ( 1 - e^{-E_{\rm 21cm}/{kT_s}} ) \sigma_{\nu}n_0
\end{eqnarray}
where $\sigma_\nu$ is the local absorption cross-section and $n_0$, 
as mentioned earlier, is the number of Hydrogen atoms in singlet 
states. Thus 21cm brightness temperature, $T_{21}$ depends on the 
difference between the neutral Hydrogen spin temperature and the CMB
temperature as well as on the optical depth of the medium  and 21cm absorption line suggests $T_s < T_\gamma$ 
(the background temperature is taken to be the CMB temperature).  
Therefore 21cm cosmology is an effective and important tool 
to understand the behaviour and evolution of cosmos in general and  
the``Dark ages" leading to the reionisation era in particular, the epoch 
from which not many cosmological or astrophysical events 
are available for probing the Universe.

The EDGES (Experiment to Detect the Global Epoch of Reionization Signature)
experiment measured and reported a strong 21cm absorption line at the 
period of cosmic dawn in the redshift range 
$14 < z < 20$. 
The trough of the absorption profile centered at 78 MHz (21cm absorption 
line redshifted to 78 MHz) in the sky averaged spectrum which corresponds
to $z = 17.2$ and at $z = 17.2$, the EDGES reported a 21cm brightness 
temperature of $T_{21} = -500^{+200}_{-500}$ mK with 99\% confidence 
limit (C.L.) \cite{EDGES}. From standard cosmology the expected 21cm brightness temperature at this redshift
is not below $T_{21} \simeq -0.2$ K and thus the observed $T_{21}$ by 
EDGES is 3.8$\sigma$ below what is expected. Since $T_{21}$ depends on 
$T_s - T_\gamma$, this observed additional cooling could be realised 
by either enhancing the background temperature $T_\gamma$ or by lowering 
the matter temperature $T_m$ ($=T_s$ at that epoch). Dark Matter interactions 
can influence both the options. The process of annihilation or
decay of Dark Matter can inject more energy into the background resulting in 
the rise of the background temperature \cite{Li_26} - \cite{Li_31}. Again it may be possible that 
Dark Matter-baryon interaction cools the baryons as the baryons interact 
with colder Dark Matter and consequently the spin temperature also goes down
as $T_s$ couples to $T_m$ at that epoch \cite{Li_32} - \cite{Li_38}. There are other ideas 
that can induce the observed larger than expected difference 
between the background temperature and $T_s$ such as axions \cite{Li_39, Li_40}, modification
of CMB temperature by Dark Energy \cite{Li_41}, interacting Dark Energy (IDE) \cite{Li_42}-\cite{Li_44}. 
The IDE models and its influence of 21cm brightness temperature has been 
discussed in Li et al \cite{Li}. In literature there exist several other attempts to explain this observational inconsistency of EDGES experiment \cite{Li_5} - \cite{Li_25}.  

Since till date no experimental observations or theoretical considerations could rule out the possible Dark
Matter - Dark Energy interaction, it would be wise to consider a non-minimal coupling between the two dark sector
components (Dark Matter and Dark Energy) instead
of treating them independently. In literature various authors
observed the interacting Dark Energy (IDE) scenario with different forms of interactions between Dark Matter and Dark Energy \cite{different_IDE} - \cite{IDE2} and discussed different phenomenological problems with such IDE models. For example the Dark Matter - Dark Energy interaction had been taken into considerations to address the coincidence problem of cosmological constant \cite{coincidence1, coincidence2}, Large Scale Structure formation \cite{LSS1, LSS2}, Hubble tension \cite{Htension1}-\cite{Htension2} etc.

In this work we consider the Dark Matter-baryon scattering and 
Dark Matter-Dark Energy interaction to evaluate the brightness temperature.
The approach is to solve six coupled differential equations that 
involves the evolution equations of Dark Matter temperature $T_\chi$, baryon 
temperature $T_b$ ($\equiv T_m$, matter temperature) , the drag $V_{b\chi}$ on the baryon velocity due to 
it's collision with Dark Matter, the evolution of electron fraction $x_e$ 
as well as the evolution of Dark Energy 
and Dark Matter densities due to the Dark Matter-Dark Energy interaction.
Note that Dark Matter-Dark Energy interaction involves energy transfer 
that affects the evolution of the Universe. The background evolution, 
if modified, would change the evolution of Hubble parameter $H(z)$. This 
would affect the optical depth and the evolution of baryon temperature and consequently affects the 21cm brightness
temperature $T_{21}$ as is evident from Eq. \ref{T21_1} above.
We have taken all these 
into account in the present work and show that Dark Matter-baryon interaction
along with a possible Dark Matter-Dark Energy interaction can well explain 
the observed additional trough in 21cm brightness temperature at cosmic dawn era 
by EDGES. In these analyses we have also aimed to provide constraints on the model parameters of a class of phenomenological IDE models by considering the results of EDGES observation. These constraints are further compared with the limits obtained from other cosmological experiments.

The paper is organised as follows. In Sect. \ref{dm-bar}, we describe the interaction between Dark Matter and baryon fluid by considering a temperature difference and a velocity difference between these two fluids. Sect. \ref{dm-de} is devoted to investigate the interaction between Dark Matter and Dark Energy while three IDE models are taken into account. In Sect \ref{temperature} the six coupled differential equations (previously mentioned in the introduction) which have to be solved to obtain 21cm brightness temperature, are described in detail. Sect. \ref{result} is dedicated to furnish our calculations and results. We compute the modifications of the evolution of Hubble parameter for different IDE models and calculate the 21cm absorption signal by switching on the Dark Matter-baryon and Dark Matter-Dark Energy interaction effects. We also observe the allowed parameter spaces for the model parameters consistent with the EDGES results. Finally in Sect. \ref{summary} we give a summary of our work along with a discussion. 

\section{Interactions Between Dark Matter and Baryon Fluids} \label{dm-bar}
In literature it has been observed that baryons can be cooled down by the interaction with a colder Dark Matter fluid \cite{19munoz} and hence DM-baryon scattering influence the brightness temperature of 21cm line. But in Ref. \cite{munoz} Munoz et al had discussed that not only the temperature differences between the two fluids (interacting Dark Matter and baryon) are responsible for the temperature change of the fluids but also the velocity differences between the two fluids will affect their temperature. The tendency to damp their relative velocity will have a heating effect in both of the fluids. The velocity difference between Dark Matter and baryon fluid generates from the fact that Dark Matter starts collapsing at the time of matter - radiation equality but the baryons experience it after they decoupled from photons. The Dark Matter-baryon relative velocity at kinetic decoupling is denoted as $V_{\chi b}\equiv V_\chi - V_b$, where $V_\chi$ and $V_b$ denote the velocities of Dark Matter and baryon respectively. This relative velocity redshifts away with the expansion of the Universe. Therefore the interactions between two fluids of different velocities  have two effects, one is to reduce the relative velocity between them (dragging effect) and the other is to equilibrate their temperature (heating/cooling effect). In the following we give the expressions and brief accounts related to these two effects.
\\


\noindent \underline{The Heating / Cooling Terms}

The interactions between two fluids of different temperature will heat up the colder fluid and cool down the warmer one. The heating rate in this case is proportional to the temperature difference of the fluids. However interactions between two fluids of same temperature but of different velocities can also contribute to the heating/cooling of the fluids.
In Ref. \cite{munoz}, this heating term has been evaluated and with this the rate of baryon heating is written as \cite{munoz}

\begin{eqnarray}
\frac{d Q_b}{dt} &=& \frac{2m_b \rho_\chi \sigma_0 e^{-r^2/2} (T_\chi - T_b)}{(m_\chi + m_b)^2 \sqrt{2 \pi} u_{\rm th}^3} + \frac{\rho_\chi}{\rho_m} \frac{m_\chi m_b}{m_\chi + m_b} V_{\chi b} D(V_{\chi b})\,\,,\label{heat}
\end{eqnarray}
where $\rho_\chi$ denotes the energy density of Dark Matter, $\rho_m$ is the energy density of total matter and $T_b$, $T_\chi$ are baryon temperature and Dark Matter temperature respectively. In the above $m_b$, $m_\chi$ are the masses of baryon and Dark Matter respectively. The first term on the r.h.s. of the above equation arises due to the temperature difference of the two fluids ($T_\chi - T_b$) and the second term comes due to the velocity difference between them. The drag term $D(V_{\chi b})$ is calculated in Ref. \cite{munoz} and is given as,
\begin{equation}
D(V_{\chi b}) \equiv -\frac{d V_{\chi b}}{d t} = \frac{\rho_m \sigma_0}{m_b+m_\chi}\frac{1}{V^2_{\chi b}}  F(r)\,\,, \label{DV_chib}
\end{equation}
where in the above $r \equiv V_{\chi b}/u_{\rm th}$, $u_{\rm th}^2 \equiv \frac{T_b}{m_b}+\frac{T_\chi}{m_\chi}$ and $F(r) \equiv {\rm erf}(\frac{r}{\sqrt{2}})-\sqrt{\frac{2}{\pi}}e^{-r^2/2}r$. In this calculations they have used the parametrization $\bar{\sigma} = \sigma_0 v^{-4}$ \cite{7munoz} for interaction cross section of Dark Matter and baryon fluid.
The heating rate of Dark Matter $\dot{Q}_{\chi}$ can also be obtained from the above equation by replacing $b$ with $\chi$ and vice versa.  When the modification of background evolution is considered due to the DM-DE interaction, evolution of Hubble parameter would be modified and hence the $\dot{Q}_b$ and $\dot{Q}_\chi$ would be affected. We have considered $H(z) = H_0 \sqrt{(\Omega_{b0} (1+z)^3 + \Omega_\chi(z) +\Omega_{\rm de}(z))}$ where we have obtained Dark Matter density parameter $\Omega_\chi(z)$ and Dark Energy density parameter $\Omega_{\rm de}(z)$ for each value of $z$ from Eqs. (\ref{rho_chi}) and (\ref{rho_de}). We have used this modified $H(z)$ to calculate $\dot{Q}_b$ and $\dot{Q}_\chi$ and hence include the modified background evolution effect.

\section{Interactions Between Dark Matter and Dark Energy} \label{dm-de}
As mentioned earlier, interactions between Dark Matter and Dark Energy will also have impacts on the absorption signal of 21cm line. For the standard cosmological model the evolution of Hubble parameter with redshift $z$ takes the form $H(z)=H_0 \sqrt{\Omega_{b0} (1+z)^3+\Omega_{\chi 0} (1+z)^3 + \Omega_{{\rm de}0} (1+z)^{3(1+\omega)}}$, where  $\Omega_{x0}$ represents the density parameter of the $x$ component of the Universe at $z=0$ ($x \equiv b$ (baryonic matter), $\chi$ (Dark Matter), de (Dark Energy)) and $\omega$ is the equation of state for Dark Energy. However, in an interacting Dark Matter-Dark Energy (DM-DE) scenario, energy density of Dark Matter ($\rho_{\chi}$) and Dark Energy ($\rho_{\rm de}$) are not evolving as $(1+z)^3$ and $(1+z)^{3(1+\omega)}$. Therefore, such interactions modify the expansion rate $H(z)$ of the Universe. This would affect the optical depth and spin temperature of the 21cm transition and consequently affects the brightness temperature ($T_{21}$) observations.

Considering the possible interactions between Dark Matter and Dark Energy, the continuity equations of their energy densities are given as \cite{Li},
\begin{eqnarray}
(1+z) H(z) \frac{d \rho_{\chi}}{d z}-3H(z)\rho_{\chi} & = & -\xi\,\,,\label{rho_chi}\\
(1+z) H(z) \frac{d \rho_{\rm de}}{d z} - 3 H(z) (1+ \omega) \rho_{de} & = & \xi\,\,, \label{rho_de}
\end{eqnarray}
where $\xi$ denotes the energy transfer between Dark Matter and Dark Energy due to their interactions. Different interacting Dark Energy (IDE) models can be obtained with different forms of $\xi$ \cite{different_IDE}-\cite{IDE2}. Here in our calculations, we take three simple and well studied phenomenological models where the energy transfer $\xi$ has following forms \cite{three_models}-\cite{three_models2}, 
\begin{center}
Model-I \hspace{5mm} $\xi=3 \lambda H(z) \rho_{\rm de}\,\,,$\\
Model-II \hspace{5mm} $\xi=3 \lambda H(z) \rho_{\chi}\,\,,$\\
Model-III \hspace{5mm} $\xi=3 \lambda H(z) (\rho_{\rm de} +\rho_{\chi})\,\,.$
\end{center}
In the above, $\lambda$ is the interacting parameter of DM-DE interactions. 
The Model-I, Model-II and Model-III can be written in a unified way as $\xi = 3 \lambda H(z) (p \rho_{\rm de} +(1-p)\rho_\chi)$ where $p=1$, $p=0$ and $p=0.5$ represent Model-I, Model-II and Model-III respectively. To observe different behaviour of the system when $\xi$ is proportional to the Dark Energy density or to the Dark Matter density or to the linear combination of both the densities, these three models are studied separately. 
The stability conditions for these phenomenological IDE models are given in Table \ref{stability}, \cite{81li, three_models2,81li_27,81li_28}. 
\begin{table}[H]
\centering
\begin{tabular}{|l|c|c|r|}
\hline
Model & $\xi$ & EOS of Dark Energy & constraints\\
\hline
I & 3 $\lambda H(z) \rho_{\rm de} $ & $-1<\omega<0$ & $\lambda<0$\\
\hline
I & 3 $\lambda H(z) \rho_{\rm de} $ & $\omega<-1$ & $\lambda<- 2 \omega \Omega_{\chi}$\\
\hline
II & 3 $\lambda H(z) \rho_{\chi} $ & $\omega<-1$ & $0<\lambda<-\omega/4$\\ 
\hline
III & 3 $\lambda H(z) (\rho_{\rm de} + \rho_{\chi}) $ & $\omega<-1$ & $0<\lambda<-\omega/4$\\
\hline 
\end{tabular}
\caption{Stability conditions of the IDE models}\label{stability}
\end{table}
Moreover these models of IDE are extensively studied and well constrained by using the PLANCK data, baryon acoustic oscillation (BAO)
data, Supernova Ia (SNIa) observational data etc. \cite{81li} - \cite{khurshydayan}. In Table \ref{constraints} we show the constraints on the model parameters. 
It can be noted from Table. \ref{constraints} that constraints are very tight for Model-II and Model-III. It is pointed out in \cite{82li} that these two models have significant effects on low $l$ region of CMB power spectrum and therefore are tightly constrained. With the perturbation formalism, the DM-DE interaction can be probed by the latter's possible influence on CMB spectrum. It is observed that if only Dark Energy part is considered i.e., $\xi=3 H(z) \lambda \rho_{\rm de}$, a stable curvature perturbation is obtained but when only the Dark Matter part is considered or the linear combination of two $(\rho_{\rm de} + \rho_\chi)$ are considered, the Dark Energy equation of state $\omega$ would be less than $-1$ to have a stable curvature perturbation and these two models affect the low $l$ region of CMB power spectrum. The stability of curvature perturbation with time dependent $\omega$ is discussed in Ref. \cite{10_44}.

Here we note that all the DM-DE interaction models considered in Table \ref{stability} or in Table \ref{constraints} do not include DM-baryon interaction (in fact such models available in the literature includes DM-DE interaction only). Within this framework we have studied the EDGES result while we include both the DM-DE and DM-baryon interactions. It appears in the following section that the EDGES result is best represented in the framework when Model-I is adopted. But there is indeed a need for obtaining the constraints by considering both DM-DE and DM-baryon interactions and observing the consequent effects on CMB power spectrum as well as BAO, SNIa. But this is for posterity. A discussion related to the CMB limit on DM-baryon scattering cross section is given in Ref. \cite{referee} but providing constraints on DM-baryon interaction or DM-DE interaction by considering both the interactions effects, is still for posterity.  
\begin{table}[H]
\centering
\begin{tabular}{|l|c|c|r|}
\hline
Model & $\omega$ & $\lambda$ & $H_0$\\
\hline
$3 \lambda H \rho_{\rm de}$ & $-0.9191^{+0.0222}_{-0.0839}$ & $-0.1107^{+0.085}_{-0.0506}$ & $68.18^{+1.43}_{-1.44}$\\
\hline
$3 \lambda H \rho_{\rm de}$ & $-1.088^{+0.0651}_{-0.0448}$ & $0.05219^{+0.0349}_{-0.0355}$ & $68.35^{+1.47}_{-1.46}$\\
\hline
$3 \lambda H \rho_{\chi}$ & $-1.1041^{+0.0467}_{-0.0292}$ & $0.0007127^{+0.000256}_{-0.000633}$ & $68.91^{+0.875}_{-0.997}$\\
\hline
$3 \lambda H (\rho_{\rm de}+\rho_{\chi})$ & $-1.105^{+0.0468}_{-0.0288}$ & $0.000735^{+0.000254}_{-0.000679}$ & $68.88^{+0.854}_{-0.97}$\\
\hline
\end{tabular}
\caption{Constraints on the parameters of the IDE models from different cosmological experiments}\label{constraints}
\end{table}

In the following we will focus on few objectives. Firstly, we will investigate the effects of DM-DE interactions along with the DM-baryon interactions, on the brightness temperature of 21cm line. Secondly, we will explore whether a IDE model, which is well in agreement with the constraints given from other experiments (constraints given in Table \ref{constraints}), is also consistent with the constraints from the EDGES observations of 21cm signal. To this end we compare the constraints on three above mentioned IDE models obtained from the EDGES results with those from other experiments. We also compute the bounds on Dark Matter mass $m_\chi$, DM-DE interaction strength $\lambda$ and dark matter-baryon interaction cross section $\sigma_0$. 

\section{Temperature Evolutions and the Background Evolutions} \label{temperature}
In this Section we compute the evolutions of various temperatures 
($T_\chi$, $T_b$ etc.) including the effects of DM-baryon interaction 
as also DM-DE interactions as discussed above. The evolution equations are 
given by the coupled differential equations,
\begin{eqnarray}
\frac{d T_\chi}{d z} &=& \frac{2 T_\chi}{1+z} - \frac{2 \dot{Q}_\chi}{3 (1+z) H(z)}-\frac{1}{n_\chi}\frac{2 \xi}{3 (1+z) H(z)}\,\,, \label{T_chi}\\
\frac{d T_b}{d z} &=& \frac{2 T_b}{1+z} + \frac{\Gamma_c}{(1+z) H(z)}(T_b - T_\gamma)-\frac{2 \dot{Q}_b}{ 3 (1+z) H(z)}\,\,. \label{T_b}
\end{eqnarray}
The above equations are obtained using the formalism described in 
Sects. \ref{dm-bar} and \ref{dm-de} and Refs. \cite{19munoz, 13munoz}. Note that, in order to include the 
DM-DE interaction in the evolution equations, we, in this work, add 
a relevant term (3rd term on the r.h.s. of Eq. (\ref{T_chi})) for the evolution equation 
of $T_\chi$. The first term on the r.h.s. is due to the adiabatic 
expansion of the Universe while the second term is due to the DM-baryon
interactions. On the r.h.s. of Eq. (\ref{T_b}) the first and third terms 
are related to the adiabatic cooling of the Universe and DM-baryon interactions respectively while the second term is due to other heating/cooling effects. In the above equations $T_\gamma=2.725(1+z)$K denotes the photon temperature and $\Gamma_c=\frac{8\sigma_T a_r T_\gamma^4 x_e}{3 (1 + f_{\rm He} +x_e) m_e c}$ is the Compton interaction rate (here $\sigma_T$ and $a_r$ are the Thomson scattering cross section and the radiation constant respectively while $f_{\rm He}$ denotes the fractional abundance of He. The free electron abundance  $x_e=n_e/n_H$ and $m_e$ is the electron mass while $c$ is the speed of light). As baryon temperature depends on the free electron fraction $x_e$, the evolution of $x_e$ also needs to be simultaneously computed.
This evolution equation is given by the relation \cite{74li}
\begin{equation}
\frac{d x_e}{d z} = \frac{C_P}{(1+z) H(z)}\left(n_H A_B x_e^2 - 4 (1-x_e)B_B e^{\frac{-3 E_0}{4 T_\gamma}}\right)\,\,,
\end{equation}
where $C_P$ represents the Peebles $C$-factor \cite{23munoz}, $E_0$ is the ground state energy of Hydrogen, $A_B$ and $B_B$ are respectively the effective recombination coefficient and the effective photoionization rate to and from the excited states respectively \cite{75li, 24munoz}. 

In this work we have also considered the variation of relative velocity $V_{\chi b}$ with redshift as discussed in \cite{munoz},
\begin{equation}
\frac{d V_{\chi b}}{dz} = \frac{V_{\chi b}}{1+z}+\frac{D(V_{\chi b})}{(1+z) H(z)}\,\,, \label{V_chib}
\end{equation}
where the first term on the r.h.s. signifies that relative velocity redshifts away with the expansion of the Universe and the second term on the r.h.s. is for the dragging force on $V_{\chi b}$.

Note that since Eqs. (\ref{heat} - \ref{V_chib}) are all coupled, DM-DE interaction
term in Eq. (\ref{T_chi}) will affect the various observables to be computed by 
solving these equations. For example, DM-DE interactions modify the 
expansion rate $H(z)$ of the Universe and evolutions of energy densities of Dark Matter $\rho_\chi$ and Dark Energy $\rho_{\rm de}$. This 
in turn modifies the terms in Eqs. (\ref{heat} - \ref{V_chib}). 
Hence the expansion history and evolutions of temperatures are coupled as well. Therefore the six equations Eqs. (\ref{rho_chi} - \ref{V_chib}) are 
to be solved simultaneously in order to obtain the 
temperature evolutions of baryons and Dark Matter where these possible  
considerations (effects of DM - baryon interaction and DM - DE interaction) are taken into account.

As mentioned in Sect. \ref{intro}, the brightness temperature of 21cm line, $T_{21}$
can be evaluated from the equation, 

 
\begin{equation}
T_{21}=\frac{T_s-T_\gamma}{1+z}(1-\exp^{-\tau})\approx\frac{T_s-T_\gamma}{1+z} \tau \,\,. \label{T_21}
\end{equation}
Thus $T_{21}$ is proportional to the difference between the spin temperature $T_s$ and the background radiation temperature $T_\gamma$. At the time of cosmic dawn due to the Wouthuysen - Field effect induced by the Ly$\alpha$ photons, $T_s \simeq T_b$ i.e., spin temperature is approximately equal to the gas temperature or baryon temperature \cite{76li} - \cite{78li}. In equation 
(\ref{T_21}), $\tau$ denotes the optical depth of the intergalactic medium and expressed as $\tau=\frac{3}{32\pi}\frac{T_*}{T_{\rm s}}n_{\rm HI}\lambda_{21}^3 \frac{A_{10}}{H(z)}$ \cite{J.R. Protchard}, where $A_{10}$ denotes the Einstein coefficient \cite{75li} for spontaneous 
downward transition from triplet to singlet state, $\lambda_{21}$ is the wavelength of 21cm line, $n_{\rm HI}$ denotes the number density of the neutral Hydrogen while $T_*$ is the temperature corresponding to the energy of the 21cm photon transition. It may be noted that the optical depth of the gas is dependent on $H(z)$ and hence will be influenced by the DM-DE interactions. For smaller values of $H(z)$, one obtains larger values of $\tau$ which give stronger absorption signal of 21cm line. 

\section{Calculations and Results} \label{result}
In this work, we simultaneously consider the Dark Matter-baryon interactions 
and Dark Matter-Dark Energy interactions and study its impact 
on the 21cm line in general and the excess trough of 21cm absorption 
line from the era of cosmic dawn reported by EDGES experiment in particular. 
We simultaneously solve the Eqs. (\ref{DV_chib} - \ref{V_chib}) by developing a computer code  to calculate and obtain the behaviours of several quantities related  to 21 cm brightness temperature $T_{21}$. Here it can be mentioned that the results are parameterised in terms of a dimensionless cross section $\sigma_{41}$, defined by $\sigma_{41}=\frac{\sigma_0}{10^{-41} {\rm cm^2}}$.

In this work velocity dependent cross section $\bar{\sigma} = \sigma_0 v^{-4}$ (here and what follows $v$ is assumed to be in units of the velocity of light $c$, if not otherwise mentioned) has been adopted. The velocity dependence of the DM - nucleon cross section or DM - baryon cross section has been addressed earlier in Ref. \cite{JCAP}. The power law of the form $\bar{\sigma} = \sigma_0 v^{n}$ is taken and several values of the index $n$ are considered and their significance for different physical DM processes are discussed, for example $n = -4$ (i.e., $\bar{\sigma} = \sigma_0 v^{-4}$) is motivated for millicharged DM \cite{17JCAP, 18JCAP}, $n=+2, -2$ for DM with electric or magnetic dipole moment and $n=-1,0,1,2....$ etc. are attributed to the case when scattering occurs in the presence of Yukawa potential \cite{23JCAP}. Recently this has also been suggested in the context of EDGES result of cooler 21cm spectrum during the cosmic dawn that the scattering of baryons by DM with a cross section of the form $\bar{\sigma} = \sigma_0 v^{-4}$ could affect the EDGES observation effectively \cite{Ranna}. Similar proposition ($\bar{\sigma} \propto v^{-4}$) in relation to EDGES 21cm result has been mentioned in \cite{JCAP}. The velocity dependent cross section ($\bar{\sigma} = \sigma_0 v^{-4}$) has also been studied in other possible DM related phenomena such as millicharge DM, hadronically interacting DM, the BAO signal and other cosmological constraints. In the present work DM - baryon cross section is taken to be $\bar{\sigma} = \sigma_0 v^{-4}$. Now the part $\sigma_0$ can in principle be calculated by adopting a particle DM model and evaluating a t-channel Feynman diagram relevant for DM - nucleon scattering. But this involves model dependent couplings for the interaction between dark sector and visible Standard Model sector. In this work we adopted a value of $\sigma_0$ of the order of $10^{-41}$ cm$^2$ which is in the laboratory experiment limit of DM - nucleus scattering. 

\begin{figure}
\includegraphics[scale=0.5]{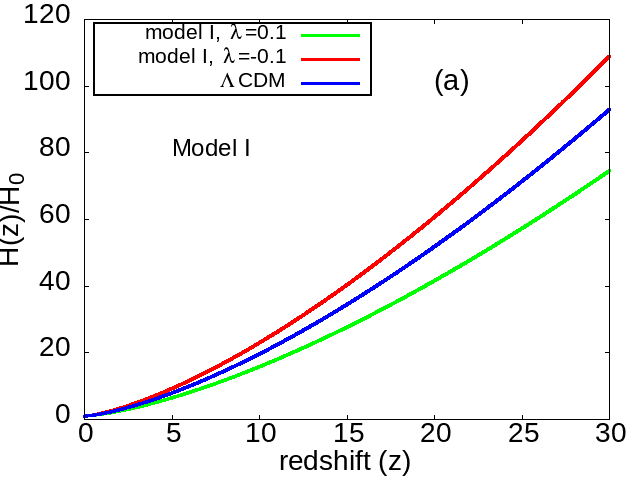} 
\includegraphics[scale=0.5]{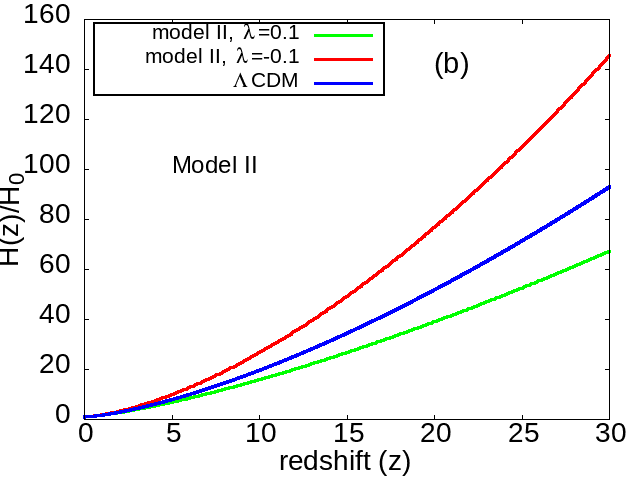}
\begin{center}
\includegraphics[scale=0.5]{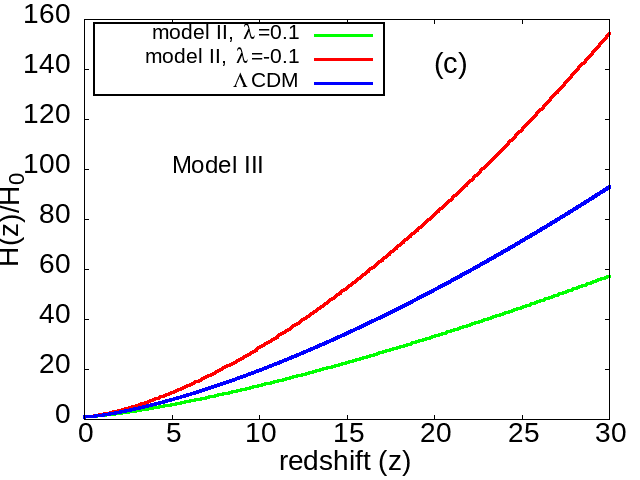}
\end{center}
\caption{Evolutions of Hubble parameter with redshift $z$ for different IDE models (Model-I in (a), Model-II in (b), Model-III in (c)) with the values of $\lambda$=-0.1, 0.1. In all the cases the evolutions are also compared with the Hubble parameter of $\Lambda$CDM (the blue line).}\label{hubble}
\end{figure}

In Fig. \ref{hubble} evolutions of Hubble parameter $H(z)$ with redshift $z$  are plotted (shown is the dimensionless quantity $H(z)/H_0$, where $H_0$ is the Hubble parameter at $z=0$) for different values of DM-DE interaction parameter $
\lambda$, namely $\lambda=-0.1,0.1$. In Fig. \ref{hubble}(a) the variations are shown for Model-I (Sect. \ref{dm-de}) while in Fig. \ref{hubble}(b) it is for Model-II (Sect. \ref{dm-de}) and in Fig. \ref{hubble}(c) it is plotted for Model-III (Sect. \ref{dm-de}). In Fig. \ref{hubble} the evolutions are also compared in all the cases with the $H(z)$ of $\Lambda$CDM model (for $\Lambda$CDM,  $\lambda=0$). These differences of the evolution of Hubble parameter $H(z)$ for different values of $\lambda$ and different models would affect the optical depth and baryon temperature or spin temperature and hence will affect the brightness temperature of 21cm line. It may be mentioned that only DM-DE interactions are considered for evaluating the results of Fig. \ref{hubble}. The effects of DM-baryon scattering along with the DM-DE interaction are discussed later. 
Such modifications of Hubble parameter due to the interaction between Dark Matter and Dark Energy are studied in literature 
and few attempts are made to describe different cosmological problems like Hubble tension \cite{Htension1} - \cite{Htension2}, the coincidence problem of cosmological constant \cite{coincidence1, coincidence2} etc with such scenarios. From Fig. \ref{hubble} we calculate the discripancy $\chi$ between the $\Lambda$CDM model and the IDE model where 
$\chi = \frac{H(z) {\rm \,\,for} \Lambda{\rm CDM \,\, model} - H(z) {\rm \,\, for \,\, considered \,\, IDE \,\, model}}{H(z) {\rm \,\, for} \Lambda{\rm CDM \,\,  model}}$ for different redshifts ($z$) and observe that it is almost constant for different redshifts. From Fig. \ref{hubble}(a) it is found that for $\lambda=0.1$, $\chi = 0.19$ at $z=10$ and $\chi \sim 0.198$ for both the values of redshift $z=600$ and $z=900$. Similar conclusions are drawn from Fig. \ref{hubble}(b) and Fig. \ref{hubble}(c). The significance of the plots in Fig. \ref{hubble} can be clarified as follows. In Fig. \ref{hubble} we observe variations of Hubble parameter for two cases namely a positive and a negative values of $\lambda$, for three different IDE models. In this case the values of $\lambda$ ($\lambda=-0.1,0.1$ in Fig. \ref{hubble}) are not necessarily within the constraints described in Table \ref{constraints} (or in Ref. \cite{81li_27})but rather they are used to demonstrate the evolution of $H(z)$ for a positive or a negative $\lambda$. But later in our calculations we use the values of $\lambda$ that respect the constraints in Table \ref{constraints} and show the variations of such models from standard $\Lambda$CDM.

\begin{figure}
\includegraphics[scale=0.5]{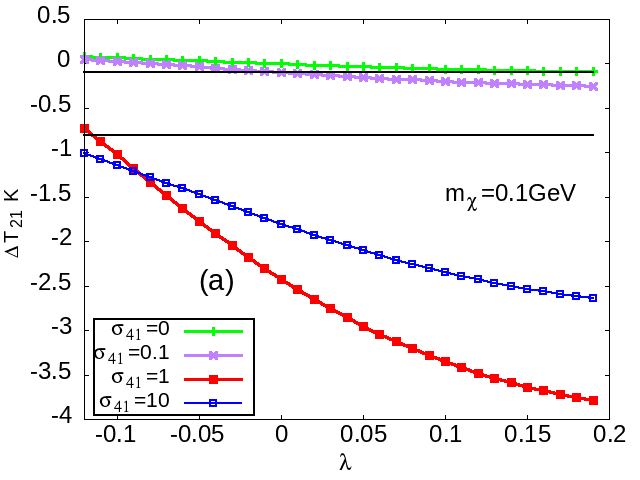} 
\includegraphics[scale=0.5]{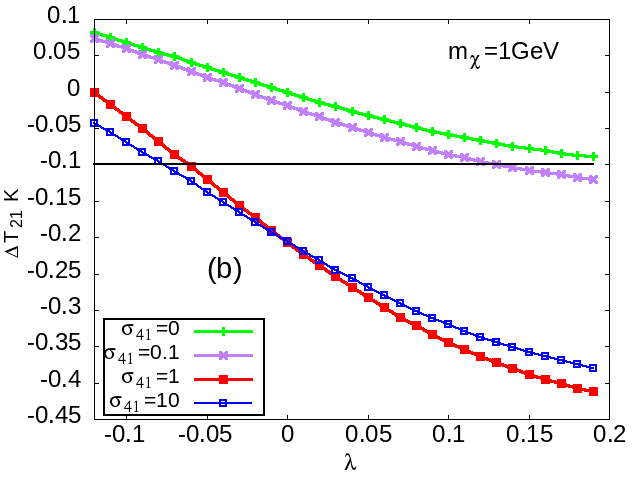}
\begin{center}
\includegraphics[scale=0.5]{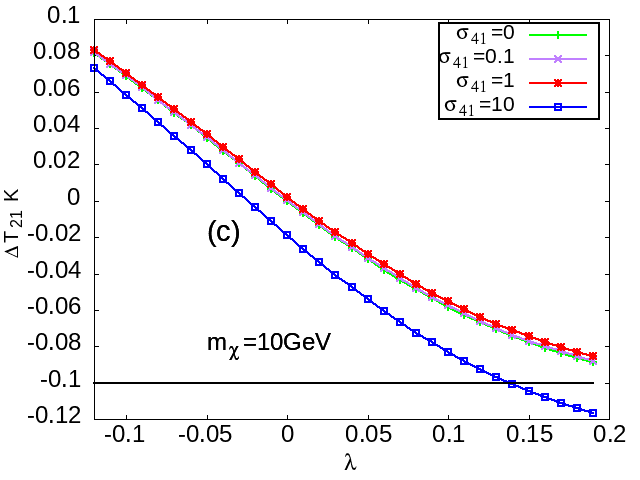}
\end{center}
\caption{The variations of $\Delta T_{21}$ ($ =T_{21}^x - T_{21}^0$) with the DM-DE interaction strength $\lambda$ for Model-I. The variations are plotted for three different values  of Dark Matter mass ($m_\chi=0.1$ GeV (in Fig. \ref{T21_model1}(a)), 1 GeV (in Fig. \ref{T21_model1}(b)) and 10 GeV (in Fig. \ref{T21_model1}(c)). For each mass the variations are compared for different values of $\sigma_{41}$. The allowed regions for EDGES results (-0.1 K$\geq \Delta T_{21}\geq$-0.8 K) are shown by solid black lines. The region between the two black lines in Fig. \ref{T21_model1}(a) and the regions below the black line for Fig. \ref{T21_model1}(b) and \ref{T21_model1}(c) are allowed for EDGES result.} \label{T21_model1}
\end{figure}

In order to better understand the influence of the effects considered 
in this work on 21cm brightness temperature, we define a quantity
$\Delta T_{21} = T_{21}^x - T_{21}^0$, where $T_{21}^0$ is the calculated 
value of brightness temperature $T_{21}$ at $z \simeq 17.2$ for $\Lambda$CDM model and 
$T_{21}^0 \simeq -0.22$ K (the expected brightness temperature at $z \simeq 17.2$. We have also obtained the value $T^0_{21}=-0.22$ K for the $\Lambda$CDM model from the framework presented in this work when $\lambda=0$, $\sigma_{41}=0$ at $z=17.2$.)  while $T_{21}^x$
corresponds to the brightness temperature at $z \simeq 17.2$ for different values 
of $\lambda$ and $\sigma_{41}$. We then 
compute the variations of $\Delta T_{21}$ with $\lambda$ for a  
set of four different 
fixed values of Dark Matter-baryon scattering cross-section $\sigma_{41}$
(in units of 10$^{-41}$ cm$^2$) namely $\sigma_{41} = 0,\,0.1,\,1,\,10$ 
while the Dark Matter mass $m_\chi$ is kept fixed at different chosen values.
Note that the choice $\sigma_{41} = 0$ signifies no Dark Matter-baryon 
interaction.
The results are plotted in three plots (plots (a), (b), (c) of Fig. \ref{T21_model1}). Fig. \ref{T21_model1}(a), 
Fig. \ref{T21_model1}(b), Fig. \ref{T21_model1}(c) correspond to the computed results with the chosen Dark Matter
masses $m_\chi = 0.1$ GeV, 1 GeV and 10 GeV respectively for Model-I in Sect. \ref{dm-de}. 
It can be observed from Fig. \ref{T21_model1}, that for all the cases,  
$\Delta T_{21}$ tends to assume more negative values as the DM-DE interaction
parameter $\lambda$ increases. This in turn signifies that the discrepancy 
in the observed value of $T_{21}$ (depth of measured trough 
of $T_{21}$ is more than expected) is better addressed for higher values of 
$\lambda$. Thus it appears that the DM-DE interaction has significant 
influence on 21cm absorption line and hence on brightness temperature at
cosmic dawn epoch. This can be envisaged from Fig. \ref{hubble} too. From Fig. \ref{hubble},
one sees that when $\lambda$ is positive, the Hubble parameter 
$H(z)$ falls below what is expected for $\Lambda$CDM (no DM-DE interaction 
case). This is also true for $z=17.2$, the redshift at which the measured value of $T_{21}$ is less than expected. Since the 
optical depth $\tau \sim 1/H(z)$, reduction in $H(z)$ 
implies increase in $\tau$ and hence one expects an enhanced negative 
brightness temperature $T_{21}$ ($T_{21} \sim 
((T_s - T_{\gamma})/(1 + z))\tau$) as observed by EDGES experiment.

It can also be noted that for every value of $m_\chi$ the discrepancy between the results expected in case of standard cosmology and EDGES results are better 
addressed for non-zero values of $\sigma_{41}$. Hence when Dark Matter baryon interactions are considered, the larger absorption signal of 21cm line can be obtained.
It is also clear from Fig. \ref{T21_model1} that smaller Dark Matter masses better obliviate the tension by producing large values of $T_{21}^x$ at $z \simeq 17.2$. But for $m_\chi=0.1$GeV and $\sigma_{41}=1, 10$ the brightness temperature are too large to fit the range of EDGES result \cite{EDGES} 
-0.1 K$\geq \Delta T_{21}\geq$-0.8 K. For other masses the results are well within the EDGES observations for positive interacting parameters and larger $\sigma_{41}$. The allowed regions for EDGES results are shown by solid black lines in Fig. \ref{T21_model1} (the region between the two black lines in Fig. \ref{T21_model1}(a) and the regions below the black line for Fig. \ref{T21_model1}(b) and \ref{T21_model1}(c)).



\begin{figure}[H]
\begin{center}
\includegraphics[scale=0.5]{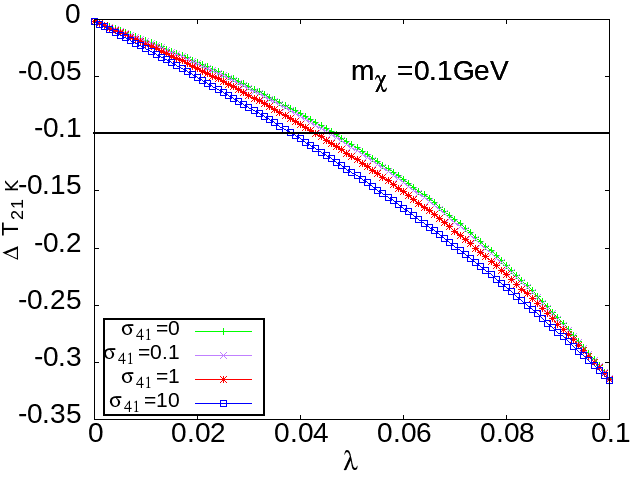} 
\caption{The variations of $\Delta T_{21}$ ($ =T_{21}^x - T_{21}^0$) with the DM-DE interaction strength $\lambda$ for Model-III. The variations are plotted for $m_\chi=0.1$ GeV and are compared for different values of $\sigma_{41}$. The regions below the black solid line are allowed for EDGES result (-0.1 K$\geq \Delta T_{21}\geq$-0.8 K).}\label{T21_model3}
\end{center}
\end{figure}

In Fig. \ref{T21_model3} we furnish similar plots as in Fig. \ref{T21_model1} but with Model-III (Sect. \ref{dm-de}). It can be observed that in this case also 21cm brightness temperature tends to have larger values for positive interaction strength and larger values of $\sigma_{41}$. In Fig. \ref{T21_model3} the region below the black solid line indicates the allowed values of $\Delta T_{21}$ from EDGES result. One may conclude from Fig. \ref{T21_model3} that the EDGES observations for $T_{21}$ are satisfied for $\lambda\gtrsim0.04$. But from Table \ref{constraints} it is found that the upper bound of $\lambda$ from other experimental constrains is $0.000989$. Hence, for this model the constraints from EDGES observations and the constraints from other experiments do not agree with each other. This is also the case when Model-II for IDE (Sect. \ref{dm-de}) is adopted in the analysis. We have also repeated our calculations even with lower DM masses but have obtained similar results. Therefore these two models do not appear to corroborate with the constraints obtained from these EDGES results.   


It may be mentioned here that in our calculations the choice of the equation of state parameter $\omega$ is made in such a way that it remains consistent with the constraints given in Table. \ref{constraints}. The initial value of relative velocity $V_{\chi b}$ is taken to be $3\times10^{-5} c$ while $V_{rms}=10^{-4} c$ \cite{21munoz}. But we have checked that even if we adopt different values 
\cite{18munoz, 22munoz} as the initial values of $V_{\chi b}$ then the average of the computed results using those initial values do not differ significantly.

It may be noted that the excess depth of the $T_{21}$ absorption signal mainly depends on DM-baryon interaction $\sigma_0$, DM-DE interaction parameter $\lambda$ and also on Dark Matter mass $m_\chi$, within the framework of the present formalism. A larger DM-baryon interaction cross section cools down baryon (and heating up Dark Matter) and thus lowers the $T_{21}$ absorption signal further. On the other hand a large positive DM-DE interaction parameter $\lambda$ lowers the Hubble expansion rate $H(z)$ which in turn increases the optical depth and hence decreases the baryon temperature. Therefore larger values of $\lambda$ are also useful for deeper $T_{21}$ absorption trough as observed by EDGES. Also note that smaller values of Dar Matter mass $m_\chi$ would lead to higher DM-baryon interaction rate thus lowering the baryon temperature.




\begin{figure}[H]
\includegraphics[scale=0.44]{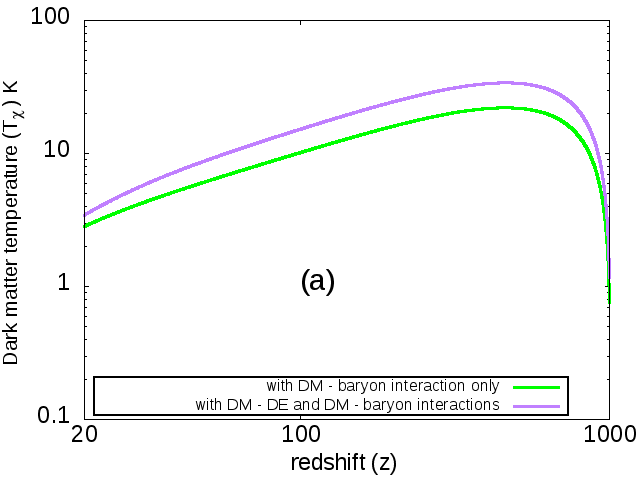} 
\includegraphics[scale=0.44]{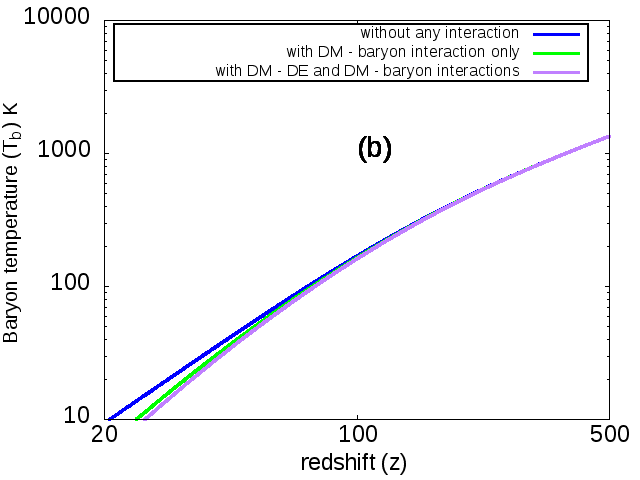}
\begin{center}
\includegraphics[scale=0.44]{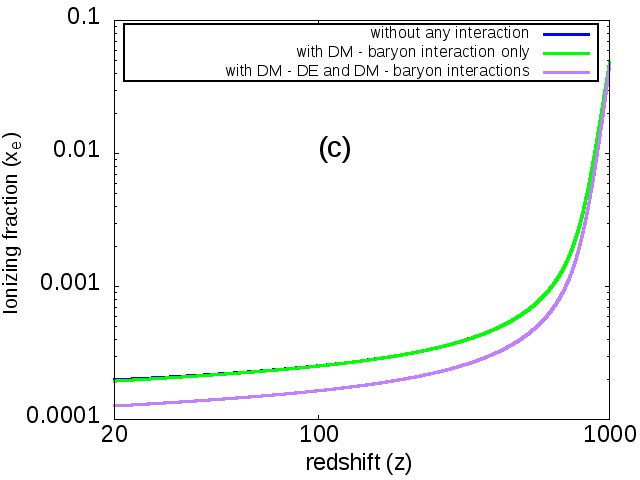}
\end{center}
\caption{The variations of Dark Matter temperature ($T_\chi$), baryon temperature ($T_b$) and ionization fraction ($x_e$) are plotted in Fig. \ref{z_vary}(a), Fig. \ref{z_vary}(b) and Fig. \ref{z_vary}(c) respectively. Results are compared for three cases say when there are no interactions between Dark Matter - baryon and Dark Matter - Dark Energy (without any interaction), when DM - baryon interaction is present but DM - DE interaction is absent (with DM - baryon interaction only) and when DM - DE interaction is present along with DM - baryon interaction (with DM - DE and DM - baryon interactions). For these plots we consider Model-I ($\xi=3 \lambda H(z) \rho_{\rm de}$) with $\lambda=0.08$, $\sigma_{41} =1$ and $m_\chi =1$ GeV.}\label{z_vary}
\end{figure}

In Fig. \ref{z_vary} the variations of Dark Matter temperature ($T_\chi$) with $z$ (in Fig. \ref{z_vary}(a)), the variations of baryon temperature ($T_b$) with $z$ (in Fig. \ref{z_vary}(b)) and the evolutions of ionization fraction ($x_e$) with $z$ (in Fig. \ref{z_vary}(c)) are plotted. We compare our results for three cases, (i) when there are no interactions between Dark Matter - baryon and Dark Matter - Dark Energy (shown by the blue coloured lines in Fig. \ref{z_vary}(b) and Fig. \ref{z_vary}(c), (ii) when DM - baryon interaction is present but DM - DE interaction is absent (the green lines in the plots of \ref{z_vary}(a) to \ref{z_vary}(c))
and (iii) when DM - DE interaction is present along with DM - baryon interaction (purple lines in the plots of \ref{z_vary}(a) to \ref{z_vary}(c)). From Fig. \ref{z_vary}(a) it can be seen that the Dark Matter temperature $T_\chi$ is increased when DM-DE interactions are taken into account. This indicates that in this scenario of DM - DE interaction, Dark Energy transfers energy to Dark Matter. From Fig. \ref{z_vary}(b) it can be noted again that the baryon temperature $T_b$ decreases when DM - baryon interaction is considered (green line) but $T_b$ decreases further when DM -DE interaction is also taken into account (purple line). Finally from Fig. \ref{z_vary}(c) one sees that the ionization fraction $x_e$ is decreased due to the presence of DM - DE interaction. For these plots we consider Model-I ($\xi=3 \lambda H(z) \rho_{\rm de}$) with $\lambda=0.08$, $\sigma_{41} =1$ and $m_\chi =1$ GeV.  

\begin{figure}[H]
\begin{center}
\includegraphics[scale=0.6]{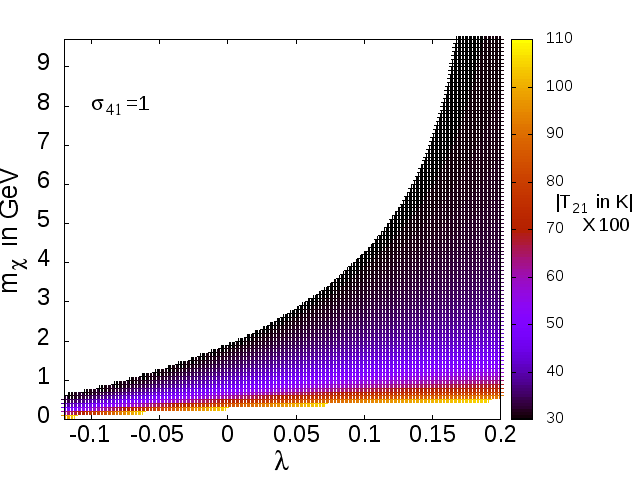}
\caption{Allowed parameter space for Dark Matter mass and DM-DE interaction strength to obey the EDGES observations.}\label{mx_lambda}
\end{center}
\end{figure}

We now calculate the allowed ranges of the parameters $\lambda$, $m_\chi$ and $\sigma_{41}$ for Model-I (Sect. \ref{dm-de}) for which the 21cm brightness temperature would satisfy the EDGES observations. In Fig. \ref{mx_lambda} we plot the allowed range of parameters in $\lambda - m_\chi$ parameter space for which the EDGES result 
-0.1 K$\geq \Delta T_{21}\geq$-0.8 K 
is satisfied. The allowed parameter space is shown by the shaded region in Fig. \ref{mx_lambda}. The parameter space on the left of the shaded region is disallowed by EDGES result. It can be seen from Fig. \ref{mx_lambda} that for larger Dark Matter mass one requires larger $\lambda$ to interpret the EDGES result. This can also be seen from Fig. \ref{mx_lambda} that the positive values of $\lambda$ are more favourable but in case $\lambda$ is negative, the Dark Matter mass has to be less than 2 GeV for obtaining the EDGES result. Also more negative the value of $\lambda$ is lighter the Dark Matter mass should be for the EDGES range of $T_{21}$ to be satisfied. For example if $\lambda \simeq -0.1$ the Dark Matter mass should be less than 1 GeV. This also corroborates the result shown in Fig. \ref{T21_model1}. Although the computations of Fig. \ref{mx_lambda} are made for $\sigma_{41}=1$, we also checked that similar allowed regions are obtained when $\sigma_{41}=0.1$ or $\sigma_{41}=10$ is considered.


\begin{figure}[H]
\includegraphics[scale=0.5]{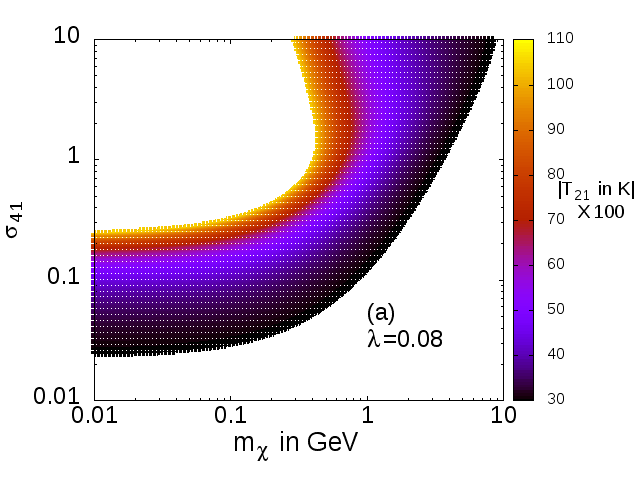}
\includegraphics[scale=0.5]{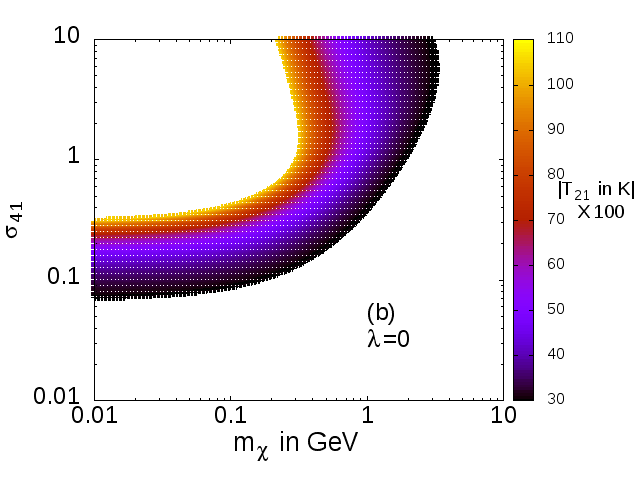}
\caption{Allowed parameter space for Dark Matter-baryon interaction cross section and  Dark Matter mass to obey the EDGES results. In Fig. \ref{sigma_mx}(a) and Fig. \ref{sigma_mx}(b) the allowed region with DM-DE interactions ($\lambda = 0.08$) and without DM-DE interactions ($\lambda=0$) are compared.}\label{sigma_mx}
\end{figure}

In Fig. \ref{sigma_mx}, the allowed regions for Model-I (Sect. \ref{dm-de}) in the $m_\chi - \sigma_{41}$ parameter 
space (that satisfies the EDGES limit 
-0.1 K$\geq \Delta T_{21}\geq$-0.8 K 
are shown while the DM-DE interaction are kept fixed at a 
chosen value. The allowed region in Fig. \ref{sigma_mx}(a) is for $\lambda = 0.08$ 
while that in Fig. \ref{sigma_mx}(b) is obtained using $\lambda = 0$ (no DM-DE interaction).
In both the cases the allowed parameter space is shown as the shaded regions 
(in Fig. \ref{sigma_mx}(a) and Fig. \ref{sigma_mx}(b)). 
Thus Fig. \ref{sigma_mx} compares the allowed $m_\chi - \sigma_{41}$ parameter 
regions with and without DM-DE interactions. It may be mentioned here that 
we have also made our calculations with other fixed values of $\lambda$ to 
obtain similar results and those are not shown. Comparing Fig. \ref{sigma_mx}(a) 
with Fig. \ref{sigma_mx}(b) it is clear that DM-DE interaction modifies the allowed
$m_\chi - \sigma_{41}$ parameter space. In fact the region of allowed parameter 
space increases when the DM-DE interaction is turned on. Therefore, DM-DE interaction increases the possibility that higher range of dark matter
mass and scattering cross-section can impart the cooling effect on baryon 
temperature, required for lowering of $T_{21}$ brightness temperature 
than what is normally expected from $\Lambda$CDM model. Also since the 
nature and mass of dark matter is still unknown, the DM-DE interaction 
effects raise the possibility of probing larger mass ranges of Dark Matter 
to have caused the cooling effect of $T_b$. Therefore DM-DE interaction
not only adds to the lowering of $T_b$ but allows to probe Dark Matter 
of different mass ranges that could have caused the observed EDGES result 
for $T_{21}$ brightness temperature.

Here we can compare Fig. \ref{sigma_mx}(b) of our work and Fig. 3 of Ref. \cite{Ranna} as both the plots are showing $T_{21}$ at $z \sim 17.2$ (regardless of astrophysics) as a function of DM mass and DM - baryon cross section when DM - DE interaction is not considered. From both the plots we can observe similar pattern of the allowed parameter space of $\sigma_{41}$ 
($\sigma_1$ in Ref.  \cite{Ranna} where $\sigma_1 =\sigma_{41} \times 10^{-19}$ cm$^2$) and $m_\chi$. In this work it is assumed that $\bar{\sigma} = \sigma_0 (v/c)^{-4}$ and $\sigma_{41}=\frac{\sigma_0}{10^{-41} {\rm cm}^2}$ as it is already mentioned earlier. But in Ref. \cite{Ranna} $\bar{\sigma}$ (or $\sigma(v)$) is assumed as $\bar{\sigma} = \sigma_1 (\frac{v}{1 {\rm km/s}})^{-4}$. Therefore by comparing these we obtain that $\sigma_1 = \sigma_{41} \times 10^{-19}$ cm$^2$. In Fig. 3 of Ref. \cite{Ranna}, contours corresponding to $T_{21}$ being more negative by 10$\%$, 50$\%$, 100$\%$ than the strongest possible absorption signal without any interaction with DM are shown and constraints on $\sigma_1$ and $m_\chi$ are given on that basis. For example $\sigma_1 > 1.5 \times 10^{-21} {\rm cm}^2$ when $m_\chi < 23$ GeV and $\sigma_1 > 3.6 \times 10^{-21} {\rm cm}^2$ while $m_\chi < 3.5$ GeV are obtained by considering contours of 10$\%$ and 50$\%$ respectively. In Fig. \ref{sigma_mx}(b) of our work we obtain the allowed $\sigma_{41} - m_\chi$ space for which the values of $T_{21}$ lie between the EDGES result ($T_{21}= -500_{-500}^{200}$ mK). But here we can easily observe that the values obtained for $\sigma_{41}$ and $m_\chi$ respect the constraints obtain in Fig. 3 of Ref. \cite{Ranna}. For example, in our case we find that to obey the EDGES result, $\sigma_{41} > 5 \times 10^{-2}$ (or $\sigma_1 > 5 \times 10^{-21} {\rm cm}^2$) and $m_\chi$ is less than 3.24 GeV. 

We have also repeated our calculations with DM-DE interactions of Model-II
and Model-III. But unlike Model-I which has been discussed above, the computed
values of $T_{21}$ fail to satisfy the EDGES results along with the other observational results for Model-II and Model-III.

\begin{figure}[H]
\includegraphics[scale=0.5]{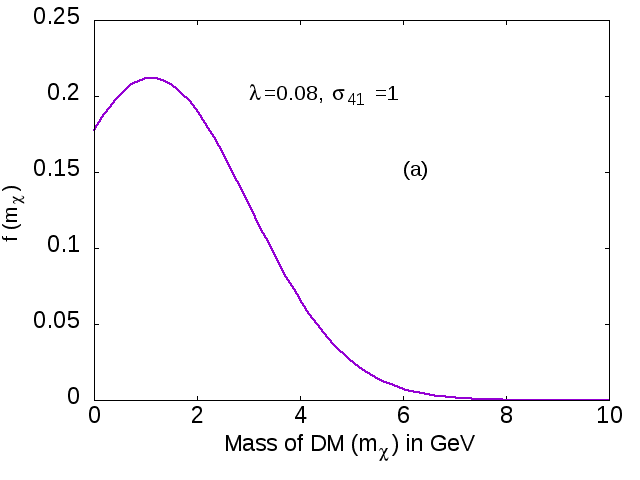}
\includegraphics[scale=0.5]{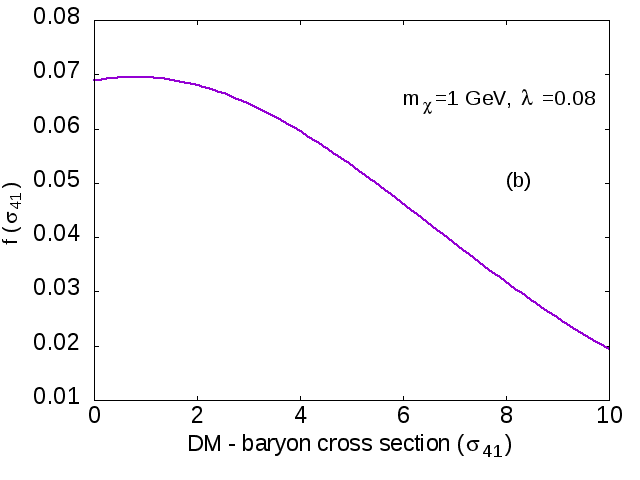}
\begin{center}
\includegraphics[scale=0.5]{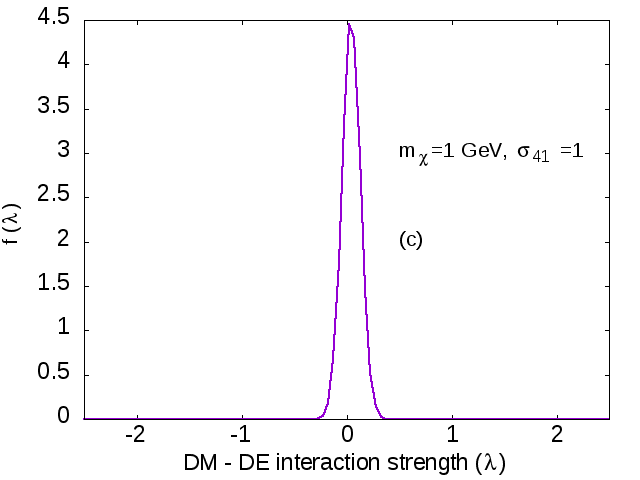}
\end{center}
\caption{Posterior distribution of mass of DM ($m_\chi$), DM - baryon cross section ($\sigma_{41}$) and DM - DE interaction strength ($\lambda$) are plotted in Fig. \ref{posterior}(a), \ref{posterior}(b) and \ref{posterior}(c) respectively. In Fig. \ref{posterior}(a) we consider fixed values of $\lambda$ and $\sigma_{41}$ at $\lambda=0.08$ and $\sigma_{41}=1$. In Fig. \ref{posterior}(b) the fixed values are taken to be $m_\chi$ = 1 GeV and $\lambda$= 0.08 while in Fig. \ref{posterior}(c) those are $m_\chi$ = 1 GeV and $\sigma_{41} =1$.}\label{posterior}
\end{figure}

Posterior distribution of mass of DM ($m_\chi$), DM - baryon cross section ($\sigma_{41}$) and DM - DE interaction strength ($\lambda$) are plotted in Fig. \ref{posterior}(a), \ref{posterior}(b) and \ref{posterior}(c) respectively. In Fig. \ref{posterior}(a) we consider fixed values of $\lambda$ and $\sigma_{41}$ at $\lambda=0.08$ and $\sigma_{41}=1$. In the distribution the mean value of $m_\chi$ ($\approx 1.12$) is corresponding to the value $T_{21} = -0.5$ K. In Fig. \ref{posterior}(b) the fixed values are taken to be $m_\chi$ = 1 GeV and $\lambda$= 0.08 while in Fig. \ref{posterior}(c) those are $m_\chi$ = 1 GeV and $\sigma_{41} =1$. The mean values of Fig. \ref{posterior}(b) and Fig. \ref{posterior}(c) are $\sigma_{41} \approx 0.795$ and $\lambda \approx 0.045$ respectively.

\section{Summary and Discussions} \label{summary}
In this work we study the effect of interacting Dark Energy with Dark Matter as well as the scattering of baryon matter with Dark Matter to explain the excess of 21cm absorption temperature reported by EDGES experiment. The EDGES experiment has observed an excess absorption signal of 21cm brightness temperature $T_{21}$  at the era of the cosmic dawn of the Universe. According to the EDGES report $T_{21}= -500_{-500}^{200}$ mK with 99\% confidence limit at redshift $z \simeq 17.2$. Expected 21cm brightness temperature from standard $\Lambda$CDM model should not be below $-0.2$ K at that redshift. Therefore (as $T_{21} \simeq \frac{T_s - T_\gamma}{1+z} \tau$) modifications in the background temperature ($T_\gamma$) or in the spin temperature ($T_s$) or in the optical depth ($\tau$) or simultaneous modifications of all of them are needed to explain the EDGES result in the cosmic dawn era. In this work we consider Dark Matter-baryon scattering as well as Dark Matter-Dark Energy interaction to evaluate the brightness temperature $T_{21}$. The approach is to solve six coupled differential equations that includes the evolutions of the energy density of Dark Matter $\rho_\chi$, energy density of Dark Energy $\rho_{\rm de}$, baryon temperature $T_b$, Dark Matter temperature $T_\chi$, free electron fraction $x_e$ and the relative velocity $V_{\chi b}$ with redshift $z$.

We note that DM-DE interaction modifies the expansion rate $H(z)$ of the Universe. This modification in the evolution of the Hubble parameter $H(z)$ would affect the optical depth of the Hydrogen cloud and the baryon temperature $T_b$ ($T_b=T_s$ at cosmic dawn epoch) and thus consequently alter the temperature of 21cm line $T_{21}$. On the other hand the scattering between the Dark Matter and baryon would also affect the baryon temperature and hence modify the $T_{21}$ signal. The Dark Matter-baryon interactions in this work include the effects of temperature differences as well as a velocity difference between these two fluids. By taking all these into account we show that Dark Matter-baryon interaction along with Dark Matter-Dark Energy interaction can well explain the observed trough of 21cm brightness temperature at cosmic dawn era by EDGES experiment. For the DM-DE interaction we adopt three interacting Dark Energy (IDE) models. We  provide constraints on the IDE models described in Sect. \ref{dm-de} on the basis of the EDGES result. Furthermore comparing these constraints with the constraints obtained from other cosmological observations (Table. \ref{constraints}) it can be noted that while the constraints obtained from EDGES observations are consistent with other experimental results for Model-I ($\xi=3 \lambda H(z) \rho_{\rm de}$), similar constraints for IDE using EDGES results when Model-II ($\xi= 3 \lambda H(z) \rho_\chi$) or Model-III ($\xi=3 \lambda H(z) (\rho_{\rm de}+\rho_\chi)$)is used for IDE are not in agreement with the same from other experimental results. Hence the last two models are in severe tension in case DM-DE interaction influences the 21cm temperature at cosmic dawn epoch.

We explore the allowed range of parameters in $\lambda - m_\chi$ and $m_\chi - \sigma_{41}$ parameter space for which the EDGES result 
-0.1 K$\geq \Delta T_{21}\geq$-0.8 K
is satisfied. It is found from our analyses that larger DM-baryon interaction cross section $\sigma_0$, larger DM-DE interaction parameter $\lambda$ and smaller DM mass $m_\chi$ are more favourable to achieve observed excess absorption feature of 21cm temperature in the cosmic dawn epoch. This can be understood by analysing Fig. \ref{T21_model1}, Fig. \ref{mx_lambda} and Fig. \ref{sigma_mx}. One can notice from the allowed regions of the plots in Fig. \ref{sigma_mx} that the presence of the DM-DE interaction enhances the possibility that higher range of DM mass and scattering cross section can lower the $T_{21}$ within EDGES range. Hence DM-DE interaction raises the possibility of probing larger mass ranges
of Dark Matter that could have influenced the cooling effects of $T_b$ . Therefore DM-DE interaction not only lower the brightness temperature $T_{21}$ but allows to probe DM of different mass ranges that could
have given rise to the observed EDGES result for $T_{21}$ brightness temperature.

One can compare the results in Fig. \ref{sigma_mx} and complementary bounds on direct detection experiments of DM. The direct detection experiments all over the world provide upper bounds on WIMP (Weakly Interacting Massive Particle) - nucleon cross sections. This can be of two types namely spin-independent and spin-dependent cross sections and in this case the DM is considered to be a WIMP which may scatter off of the nucleus of the detecting material and thereby the nucleus suffers a recoil with a certain recoil energy. As different experiments use different detecting materials (and hence different nucleus) the DM - nucleons scattering cross section bounds are provided instead of DM - nucleus scattering. The DM-nucleus scattering is given by the general expression $\sigma =\frac{4\mu^2 f_p^2 A^2}{\pi}$ 
, where $f_p$, $\mu$ and $A$ denote proton form factor, DM-nucleon reduced mass and mass number of the nucleus of the detecting material respectively. But the DM-nucleon scattering cross section is given by $\sigma^n =\frac{4\mu^2_n f_n^2}{\pi}$ with $f_p \approx f_n$, where $f_n$ is the neutron form factor. Such bounds can span upto the WIMP mass of $\mathcal{O}(10^4)$ GeV but in the lighter mass side these bounds are effective upto the order of 7 GeV because of the detector constraints (threshold). But there are a few experiments that provide such bounds even below 7 GeV which could be of interest in the present work.

In the present work however the ``baryon'' is a general term for all the known matter in the Universe and here DM-baryon cross section is parametrized by $\bar{\sigma}=\sigma_{41}v^{-4}$. Fig. \ref{sigma_mx} of this manuscript suggests that the figures span for the DM mass upto the order of 10 GeV and cross section upto $\mathcal{O}(10)$ in units of $10^{-41}$ cm$^2$. The comparison of these plots with the ones given by the DM direct detection experiments are not very useful because in Fig. \ref{sigma_mx} the allowed values of DM mass goes upto about 8 GeV and not beyond. Even if a complementary bounds can be considered for SNOLAB experiment, it provides the upper bound of DM mass below the order of 8 GeV and the spin dependent cross section is around $10^{-40}$-$10^{-44}$ cm$^2$ \cite{snowlab}. But this can be in the range of our plot.

In the end we would like to comment that with more possible data expected when new experiment namely Square Kilometer Array (SKA) will be operational and with further analysis of those results we expect to gain a better understanding of the physics of 21cm absorption signature.
 \vskip 1cm
\noindent{\large \bf Acknowledgements}  

\noindent One of the authors U.M. acknowledges Council of Scientific \& Industrial Research (CSIR), Government of India for supporting her with a fellowship grant as Senior Research Fellow (SRF) with the fellowship Grant No. 09/489(0106)/2017-EMR-I and KKD acknowledges  financial  support  from BRNS through a project grant (sanction no: 57/14/10/2019-BRNS).  


\end{document}